\documentclass[amsmath,amssymb,superscriptaddress,showpacs]{revtex4}
\usepackage{graphicx}
\usepackage{dcolumn}
\usepackage{bm}
\usepackage[colorlinks=true, citecolor=blue, urlcolor = blue, linkcolor= red, bookmarks=true]{hyperref}

\begin{document}
\title[]{ Photon orbits and phase transition for gravitational decoupled Kerr anti-de Sitter black holes}

\author{Suhail Khan}\email{suhail@ctp-jamia.res.in} 
\affiliation{Centre for Theoretical Physics, 
	Jamia Millia Islamia, New Delhi 110025, India}
\author{Shafqat Ul Islam} \email{Shafphy@gmail.com}
\affiliation{Astrophysics and Cosmology Research Unit, 
	School of Mathematics, Statistics and Computer Science, 
	University of KwaZulu-Natal, Private Bag 54001, Durban 4000, South Africa}
\author{Sushant~G.~Ghosh }\email{sghosh2@jmi.ac.in}
\affiliation{Centre for Theoretical Physics, 
	Jamia Millia Islamia, New Delhi 110025, India}
\affiliation{Astrophysics and Cosmology Research Unit, 
	School of Mathematics, Statistics and Computer Science, 
	University of KwaZulu-Natal, Private Bag 54001, Durban 4000, South Africa}
\author{Sunil D. Maharaj} \email{ maharaj@ukzn.ac.za}
\affiliation{Astrophysics and Cosmology Research Unit, 
	School of Mathematics, Statistics and Computer Science, 
	University of KwaZulu-Natal, Private Bag 54001, Durban 4000, South Africa}
\begin{abstract}
Interpreting the cosmological constant as the energy of the vacuum and using a gravitational decoupling approach leads to a new Kerr–anti-de Sitter (AdS) black hole. The metric of the new Kerr-Ads is more straightforward than the standard Kerr-Ads and geometrically richer, showing the rotation's impact as a warped curvature. We investigate the relationship between the unstable photon orbits and thermodynamic phase transition to the new Kerr-Ads black hole background.  We derive an exact expression for thermodynamic properties of black holes, including mass ($M$), Hawking temperature ($T$), entropy ($S$), heat capacity ($G$), and free energy ($G$), by relating the negative cosmological constant to positive pressure through the equation $P = -\Lambda/8\pi= 3/8\pi l^2$, where $l$ represents the horizon radius, and by introducing its conjugate variable as the thermodynamic volume $V$.  When $P < P_c$, black holes with $C_P > 0$ exhibit stability against thermal fluctuations, while those with $C_P \leq 0$ are unstable. Our analysis of Gibbs free energy reveals a phase transition from small globally unstable black holes to large globally stable ones. Additionally, investigating the system's $P-V$ criticality and determining the critical exponents shows that our system shares similarities with a Van der Waals (vdW) fluid. In the reduced parameter space, we observe nonmonotonic behaviours of the photon sphere radius and the critical impact parameter when the pressure is below its critical value. It indicates that alterations in the photon sphere radius and the minimum impact parameter can act as order parameters for the phase transition between small and large black holes.
In discussing the applicability of the Maxwell equal area law, we highlight the presence of a characteristic vdW-like oscillation in the $P-V$ diagram. This oscillation, denoting the phase transition from a small black hole to a large one, can be substituted by an isobar. Furthermore, we present the distribution of critical points in parameter space and derive a fitting formula for the co-existence curve. 
\end{abstract}

\keywords{Black holes, critical phenomena, phase structure, circular orbit}


\maketitle

\section{Introduction}
The Schwarzschild black hole consistently displays negative heat capacity, implying that as the black hole absorbs energy, its temperature decreases, making it thermodynamically unstable. When considering the anti-de Sitter (AdS) space, the Hawking-Page phase transition delineates a transition between the stable large black hole and thermal gas phases \cite{Hawking:1982dh}. This transition provides insights into the thermodynamic behaviour of black holes in AdS space and contributes to our understanding of the intricate interplay between gravitational and thermodynamic phenomena \cite{Kumar:2024sdg,Rehan:2024dsg,Kumar:2024bls,Sood:2024rfr,Ghosh:2020tgy}. The Hawking-Page phase transition is interpreted as a confinement/deconfinement phase transition within gauge theory \cite{Witten:1998zw}, can be analogously understood as a solid/liquid phase transition \cite{Altamirano:2013ane}, with the cosmological constant viewed as pressure, $P = -\frac{\Lambda}{8 \pi}$ \cite{Ali:2023ppg,Kumar:2023gjt,Sood:2022fio,Ali:2019rjn}.  
 
This explanation provides a distinct viewpoint on the thermodynamic characteristics and phase dynamics of black holes in AdS space, enhancing our system comprehension.  Additionally, with this methodology, the corresponding complementary measure is the thermodynamic volume of the black hole system. By including this pressure and volume factor, the first law of black hole thermodynamics accurately corresponds to that of a conventional thermodynamic system.  In this method, the thermodynamic volume of the black hole system is the corresponding complementary quantity. This pressure and volume term makes black hole thermodynamics' first law match that of a conventional thermodynamic system.
Moreover, scientists have discovered a phase change between small and large black holes, relating it to the liquid-gas phase transition observed in a van der Waals (VdW) fluid \cite{Kubiznak:2012wp}. Other intriguing phase transitions have been discovered beyond this small-large black hole transition \cite{Gunasekaran:2012dq,Zou:2013owa,Altamirano:2013ane,Altamirano:2013uqa,Frassino:2014pha,Wei:2014hba,Dolan:2014vba,Shao-Wen:2016uv,Hennigar:2016xwd,Hendi:2015hoa,Hendi:2015bna,Hendi:2016yof,Hendi:2017fxp,Momeni:2016qfv,Chakraborty:2015hna,Wei:2015iwa}. These findings contribute to a richer understanding of black hole thermodynamics' diverse and complex phase behavior.  A significant alteration in the slope of the quasinormal modes (QNMs) both before and after the thermodynamic phase transition of the black hole was observed \cite{Liu:2014gvf}, which presents a dynamic means of investigating the phase transition and was broadened to other AdS black holes \cite{Mahapatra:2016dae,Chabab:2016cem,Zou:2017juz,Prasia:2016esx,Liang:2017ceh}. 

These findings suggest that, under low-temperature or low-pressure conditions, the phase transition can be effectively examined by monitoring changes in the QNM frequency slope. Conversely, additional measurements, such as observing nonmonotonic behaviours in the imaginary part of the QNM frequencies, are necessary under high-temperature or high-pressure conditions \cite{Liang:2017ceh}.  Moreover, it is widely acknowledged that QNMs are closely associated with unstable photon orbits \cite{Goebel:1972apj,Cho:2011sf,Mashhoon:1985prd}. Consequently, they play a crucial role in significant gravitational phenomena, including the study of shadows and lensing \cite{Bozza:2002zj,Cardoso:2008bp,Stefanov:2010xz,Wei:2011zw,Hod:2011cxc,Wei:2013mda,Raffaelli:2014ola,Cardoso:2016rao,Konoplya:2017wot}.
QNMs can be considered as the vibration frequencies associated with photon orbits, and their derivation is based on the characteristics of these orbits \cite{Dolan:2009nk}. 

Given these observations, it is logical to establish a connection between photon orbits and black hole phase transitions. The objective is to examine whether the properties of photon orbits can serve as indicators for black hole phase transitions. A proposal suggested that the presence of photon orbits might indicate a potential York-Hawking-Page phase transition \cite{Cvetic:2016bxi,Gibbons:2008ru}. Subsequently, a more explicit connection between photon orbits and phase transitions in the Reissner-Nordström-AdS black hole \cite{Wei:2017mwc} and the rotating Kerr-AdS black hole \cite{Wei:2018aqm} were elucidated. In particular, the photon orbit radius, minimum impact parameter, and temperature demonstrate non-monotonic behaviour under low-pressure conditions,
indicating a phase transition. Specifically, it has been observed that the temperature and pressure associated with the extremal points of the radius and minimal impact parameter align closely with the thermodynamically metastable curve for Kerr-AdS black holes \cite{Wei:2018aqm}. Investigation later extended to encompass other black holes \cite{NaveenaKumara:2019nnt} and gravity theories \cite{Chabab:2019kfs,Han:2018grg,Xu:2019yub,Li:2019dai,Du:2022quq}. Lately, a comprehensive confirmation of the relationships between photon orbits and phase transitions is provided \cite{Du:2022quq,Zhang:2019tzi,Sood:2024rfr,Ali:2023ppg}. 

Motivated by these considerations, our primary focus is investigating the relationship between photon orbits and phase transitions in gravitationally decoupled new Kerr-Ads black holes. The paper will refer to the original Kerr-Ads black hole as the Kerr-Ads black hole.

This paper is structured as follows: In Sec. \ref{sec2}, we compare Kerr-Ads and new Kerr-Ads black holes and briefly overview the new Kerr-Ads black holes obtained through the gravitational decoupling approach. This section also includes a detailed discussion of the horizon structure to impose constraints on parameters and explains the allowed parameter space. Further, Sec. \ref{sec3} covers the thermodynamic stability and P-V criticality of new Kerr-Ads black holes. Sec. \ref{sec4} presents an analysis of geodesics for new Kerr-AdS black holes and explores the relationship between phase transitions and photon orbits. Finally, in Sec. \ref{sec5}, we provide concluding remarks for the paper.

\section{Kerr-AdS  black holes} \label{sec2}
Before introducing the new Kerr-AdS solution, let's briefly review the standard Kerr-AdS solution discovered by Carter \cite{Carter:1973rla}.
The  Kerr-AdS black holes are solutions to the Einstein field equations that describe the gravitational field around a rotating black hole in spacetime with a negative cosmological constant.  The metric of Kerr-AdS spacetime in the Boyer-Lindquist coordinates reads ~\cite{Carter:1973rla}  
\begin{eqnarray} \label{kdsstandard}
ds^{2} &=& \left[\frac{\Delta-\Delta_\theta\,a^2\sin^{2}\theta}{{\rho^2\,\Xi^2}}\right] dt^{2}-
\frac{{\rho}^{2}}{{\Delta}}\,dr^{2}- \frac{{\rho}^{2}}{{\Delta_\theta}}\,d\theta^{2}+
 \frac{2 {a} \sin^{2}\theta}{{\rho}^{2}\;\Xi^2}\left[\Delta_\theta(r^2+a^2)-\Delta \right]
dt d\phi\nonumber \\ && -\frac{\sin^{2}\theta}{{\rho}^{2}\;\Xi^2}\left[{{\left(r^{2}+{a}^{2}\right)^{2} \Delta_\theta
-{\Delta}\, a^2\sin^{2}\theta}}\right]\;d\phi^{2} 
\end{eqnarray}
where
\begin{eqnarray}
\rho^2 = r^2+a^2 \cos^{2}\theta, ~~~~~~ \Delta = \left(r^2+a^2\right)\left(1+\frac{r^2}{l^2}\right)-2 M r,  ~~~~~~ \Delta_\theta  =  1 - \frac{a^2\cos^{2}\theta}{l^2}~~~~~ \text{and} ~~~~~~  \Xi =  1 - \frac{a^2}{l^2}. 
\end{eqnarray}
Here $M$ is the black hole mass, and $a$ is the spin parameter. The metric \eqref{kdsstandard} is an exact $\Lambda$-vacuum solution of the Einstein field equations 
\begin{eqnarray}\label{einst-L}
 R_{\mu\nu} = \frac{3}{l^2}g_{\mu\nu}.
\end{eqnarray} 
The curvature radius $l$ is related to negative cosmological constant via $\Lambda={-3}/{l^2}$. The Kerr-AdS black hole metric~\eqref{kdsstandard}, similar to the Kerr black hole, is independent of the $t$ and $\phi$ coordinates. Consequently, it possesses two killing vectors$\eta_{(t)}^{\mu}=\delta^{\mu}_t$ and $\eta_{(\phi)}^{\mu}=\delta^{\mu}_{\phi}$. Hence, the four-momentum components related to the translation along the $t$ and $\phi$ coordinates remain constant during the motion. The event horizon-null stationary surface represents the position where future-directed null geodesic rays originating from the black hole cannot propagate to extremely far distances. The event horizon ($r_{+}$) of metric (\ref{kdsstandard}) is located at the largest root of   
\begin{equation}
  \Delta(r_{+})=  \left(r_+^2+a^2\right)\left(1+\frac{r_+^2}{l^2}\right)-2 M r_+=0,
\end{equation}

Furthermore, the black hole metric given by Eq.(\ref{kdsstandard}) exhibits singular behaviour at points where $\Sigma \neq 0$ and $\Delta=0$, constituting a coordinate singularity known as the event horizon (EH). Analysis of the equation $\Delta=0$ shows that non-zero parameters $a$ and $l$ can be found, resulting in $\Delta$ having a minimum and two positive roots $r_{\pm}$.  Indeed, there exists a critical value  $l_c$ for a given $a$, and likewise a critical value  $a_c$ for a given $l$, such that it has degenerate horizons ($r_{-}=r_+=r_E$) corresponding to an extremal black hole (cf. Fig.~\ref{plot1}). When $l>l_c$ (for e.g., for $a=0.8$, we have $l_c=1.337$), $\Delta=0$  admits two positive real roots $r_{\pm}$ (with $r_{+} \geq r_{-}$) and has no zeros for  $l<l_c$. They correspond, respectively, to a non-extremal black hole with an event horizon $(r_{+})$ and a Cauchy horizon $(r_{-})$ or to either a no black hole spacetime or naked singularity.

Furthermore, we can consider observers who remain stationary (observers with zero angular momentum with spatial infinity). Yet, they experience a non-zero angular velocity $\omega = {d\phi}/{dt}$ due to the phenomenon of frame dragging. For the Kerr-AdS  black hole, the ZAMO's angular velocity is \cite{Blagojevic:2020edq}
\begin{eqnarray}
\omega &=& -\frac{g_{t{\phi}}}{g_{{\phi}{\phi}}}=-\frac{a \Xi  \left(\left(a^2+r^2\right) \Delta _{\theta }-\Delta \right)}{\left(a^2+r^2\right)^2 \Delta _{\theta }-a^2 \sin ^2\theta  \Delta},\\
\omega|_{r=r_{+}} &=& -\frac{a\Xi}{a^2+r^2_{+}}. \label{extreme}
\end{eqnarray}
As a result, an observer outside the event horizon rotates at an angular velocity $\omega$, which increases as the observer gets closer to the black hole. The observer at the horizon starts to maximally co-rotate at a speed equal to the black hole's so that the angular velocity (as measured at infinity) at every location on the horizon is the same, as given by Eq.~(\ref{extreme}). The angular velocity does not vanish for large $r$. 
Additionally, we can determine the rotational velocity by using the Killing vector's null property as 
\begin{eqnarray}
\Omega &=& -\frac{g_{t{\phi}}}{g_{{\phi}{\phi}}}\pm\sqrt{\left(\frac{g_{t{\phi}}}{g_{{\phi}{\phi}}}\right)^2-\frac{g_{tt}} {g_{{\phi}{\phi}}}}=-{\omega}\pm\sqrt{{\omega^2}-\frac{g_{tt}} {g_{{\phi}{\phi}}}},\nonumber\\
&=& \frac{\Xi  \left( \sqrt{\Delta _{\theta } \Delta} \left(a^2 \cos 2 \theta +a^2+2 r^2\right)+2 a \Delta -2 a \left(a^2+r^2\right) \Delta _{\theta }\sin\theta \right)}{2 \sin \theta  \left(\left(a^2+r^2\right)^2 \Delta _{\theta }-a^2 \sin ^2\theta \Delta \right)}.
\end{eqnarray}
At the event horizon  $\Omega_{+}=\omega_{+}$. Further, the surface gravity for the metric \eqref{kdsstandard} is given by
\begin{equation}
    \kappa=\frac{(\partial_r \Delta)_{r_{+}}}{2(r^2_+ +a^2)}=\frac{r_{\rm +}}{2(r_{\rm +}^{2}+a^{2})}
   \left(1-\frac{a^{2}}{r_{\rm +}^{2}}+3\frac{r_{\rm +}^{2}}{l^{2}}+\frac{a^2}{l^2}\right).
\end{equation}
The surface gravity $\kappa$ and angular velocity $\omega$ are defined locally on the horizon and remain constant across the horizon of any stationary black hole. This is an expanded version of the Zeroth law of black hole mechanics.

\subsection{New Kerr-AdS black holes via GD approach}
Considering the Schwarzschild–Anti-de Sitter (SAdS) solution and applying the gravitational decoupling (GD) Approach for axially symmetric systems, we construct the rotating version \cite{Ovalle:2017fgl,Ovalle:2018gic,Ovalle:2020kpd}. This leads us to a novel black hole (BH) solution within an AdS spacetime, distinct from both $\Lambda$-vacuum solutions and the Plebański-Demiański family of metrics. It is asymptotically anti-de Sitter and includes the Kerr line element as a specific instance, hence termed New Kerr-Ads black holes. Here, we review the gravitational decoupling (GD) approach, which produces spherically symmetric hairy black holes. This approach has been expanded to include rotating cases such as Kerr-Newman black holes \cite{Contreras:2021yxe}, hairy-Kerr black holes \cite{Ovalle:2020kpd}, and Kerr de Sitter (dS) black holes \cite{Ovalle:2021jzf}.

Ovalle \textit{et. al.}~\cite{Ovalle:2020kpd} (also see \cite{Ovalle:2017fgl,Ovalle:2018gic}) developed a straightforward GD Approach to generate deformed solutions to the known GR solution due to the extra sources. Thus, one obtains a methodical and simple procedure for extensions of axially symmetric black holes utilising the GD approach \cite{Contreras:2021yxe}.  We begin with the field equations of Einstein as
\begin{equation} \label{corr2}
   G_{\mu\nu} = k^2 ({T}_{\mu\nu} + S^{\rm}_{\mu\nu}),
\end{equation}
where $k^2=8\pi G_N$. The Einstein tensor is denoted by $G^{\rm}_{\mu\nu}$, the energy-momentum tensor (EMT) of the known solution is denoted by ${T}_{\mu\nu}$, and the EMT of the new source is denoted by $ S^{\rm}_{\mu\nu}$ \cite{Ovalle:2017fgl,Ovalle:2018gic}. In the Boyer-Lindquist coordinates, we consider a generic extension of the Kerr black hole as \cite{Toshmatov:2017zpr}
\begin{eqnarray}\label{kerrex}
ds^2 & = &  \left[ 1- \frac{2r\,\tilde{m}(r)}{\rho^2} \right] dt^2 -
\frac{\rho^2}{\Delta}\,dr^2 - \rho^2 d \theta^2+ \frac{4ar\,\tilde{m}(r)
 \sin^2 \theta}{\rho^2 }\,dt\; d\phi
-\left[r^2+ a^2 +
\frac{2a^2 r\, \tilde{m}(r)\sin^2 \theta}{\rho^2} 
\right]\sin^2\theta\,d\phi^2,
\end{eqnarray}
where 
\begin{equation}
  \Delta=r^2 + a^2 - 2r\,\tilde{m}(r).
\end{equation}
For spinning compact objects, such as black holes, Eq.~\ref {kerrex} can be used; this includes the well-known Kerr black holes as special examples where $\tilde{m}(r)=M$. Additionally, the well-known Schwarzschild black holes result if $a=0$ as:
\begin{eqnarray}\label{metric}
ds^2 & = &  \left[ 1- \frac{2\tilde{m}(r)}{r} \right]   dt^2 -  \left[ 1- \frac{2\tilde{m}(r)}{r} \right]^{-1}dr^2 - r^2 d\Omega^2. 
\end{eqnarray}
On using Einstein Eq.~(\ref{corr2})  with $S_{\mu\nu}=0 $, we obtain 
\begin{equation}\label{onedim}
k^2T^0_0=k^2T^1_1=\frac{2m'}{r^2},~~~~~~ k^2T^2_2=\frac{2m''}{r}.    
\end{equation}
$T_{\mu\nu}$ contains an energy density $\epsilon=T_0^0$, a radial pressure $p_r=-T^1_1$ and tangential pressure $p_t=-T^2_2=-T^3_3$.
The GD approach can yield spinning black hole spacetimes by deforming the spherically symmetric static black hole solution of GR, such as by producing nontrivial Kerr extensions or hairy Kerr black holes. Assume that the vacuum $T^{\rm}_{\mu\nu}=0$ is represented by the $\tilde{m}(r)= m(r)$ alone. Adding the other sources $S_{\mu\nu}$ results in \cite{Ovalle:2017fgl,Ovalle:2018gic}
\begin{equation}\label{mass}
\tilde{m}(r)= m(r) + m_s(r), 
\end{equation}
Thus, the $T_{\mu\nu}$ and $S_{\mu\nu}$, respectively, yield the mass functions $m$ and $m_s$. The Schwarzschild-AdS, which is produced by combining the spherically symmetric vacuum tensor $T_{\mu\nu}=0$ with the vacuum energy as $S_{\mu\nu}= -3/l^2 g_{\mu\nu}$, is a well-known and basic example of gravitational decoupling \cite{Ovalle:2017fgl,Ovalle:2018gic}. Thus from the Eq.~(\ref{onedim}) mass functions $m(r)$ and $m_s(r)$ for spherically symmetric and $\Lambda$-vacuum, respectively,  can be identified as   
\begin{equation} \label{m2}
m(r) = M_1,~~~~~~~
m_s(r)=M_2-\frac{r^3 }{2 l^2},
\end{equation}
With $M = M_1 + M_2$, where $M_1$ and $M_2$ are constants expressed in length units, we obtain the total Misnar-Sharp mass function from Eq.~\eqref{mass} as
\begin{equation}\label{mt}
\tilde{m}(r) =  M - \frac{r^3 }{2 l^2}.
\end{equation}
\begin{figure*}[t]
	\begin{centering}
		    \includegraphics[scale=0.8]{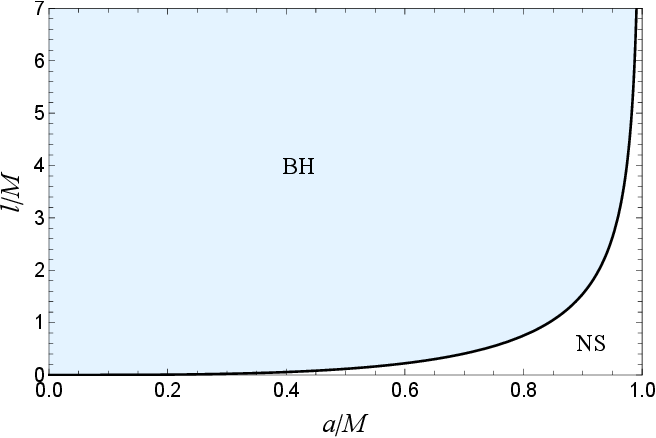}
	\end{centering}
\caption{ The parameter space $(l/M - a/M)$  of the new Kerr-AdS black hole metric is illustrated, with the dark lines representing extremal black holes that distinguish between black holes and naked singularities.}\label{plot1a}		
\end{figure*}
By plugging the Eq.~(\ref{mt}) in metric (\ref{metric}), we obtain the spherically symmetric Schwarzschild AdS  black hole surrounded by matter with conserved $S_{\mu\nu}$ satisfying the strong energy condition and  given by 
\begin{eqnarray}\label{metric3}
ds^2 & = &  \left[ 1- \frac{2M}{r} + \frac{r^2}{l^2} \right] dt^2 -  \left[ 1- \frac{2M}{r} + \frac{r^2}{l^2} \right]^{-1}dr^2 - r^2 d\Omega^2.
\end{eqnarray}

Then, the new Kerr-AdS can be easily obtained by substituting the mass function  \eqref{mt} into metric \eqref{kerrex}, which reads
\begin{eqnarray} \label{newKerrds}
ds^{2} &=& \left[\frac{\Delta-a^2\,\sin^{2}\theta}{\rho^2}\right] dt^{2}	-\frac{{\rho}^{2}}{{\Delta}}\,dr^{2}- {\rho}^{2}\,d\theta^{2}  + \frac{2\, {a}\sin^{2}\theta}{{\rho}^{2}}\left(r^2+a^2-\Delta\right) 
dt d\phi - \frac{\mathcal{A}}{{\rho}^{2}} d\phi^{2},
\end{eqnarray}
where 
\begin{eqnarray}\label{deltaN}
\mathcal{A}= [(a^2+r^2)^2-\Delta{a}^{2}\sin^{2}\theta] \sin^{2}\theta~~~~~~ \Delta  ={a}^{2}  -2 M r +r^2 + \frac{r^4}{l^2},     
\end{eqnarray}
While  Eq.~\eqref{newKerrds} appears more straightforward than Eq.~\eqref{kdsstandard} upon observation, it possesses a rich spacetime structure.  In contrast to the metric in Eq.~\eqref{newKerrds}, the Kerr-AdS metric \eqref{kdsstandard} is a $\Lambda$-vacuum solution.
The AdS geometry in nonstandard coordinates, whose metric components depend on the parameter $a$, describes the spacetime in Eq.~(\ref{newKerrds}) for $M=0$. As we find, the curvature for the line element  \eqref{newKerrds}  is given by 
\begin{equation}
    \tilde{R}_{\mu\nu}=\frac{3\,r^2 }{l^2\,\rho^2} g_{\mu\nu} \ne R_{\mu\nu} \label{R}.
\end{equation}
which is quite different from the Eq. \eqref{einst-L}. One noteworthy observation is that all other regions show warping except for the equatorial plane, where the curvature is constant. The solution \eqref{newKerrds} will hereafter be referred to as the new Kerr-AdS for clarity. The curvature effect becomes increasingly prominent close to the rotating distribution, that is, $r\sim\,a$, and eventually vanishes to the point where, for $r>>a$, $R\sim 12/l^2$. Since this effect is a constant-curvature solution by construction, it cannot occur in a Kerr-AdS spacetime. We conclude that to clarify the impacts of the spinning item in its immediate surroundings, the line element~\eqref{newKerrds} is required. Regarding the new curvature that depends upon $r$ and $\theta$, we can write  in Eq.~\eqref{R} as
\begin{equation} \label{R2}
\tilde{R}(r,\theta)=\frac{12\,r^2}{l^2\,\rho^2}=\frac{12}{l^2}\left[\frac{r^2}{r^2+{a}^{2}\cos^{2}\theta}\right]
\end{equation} 
It is evident from the preceding formula that the vacuum energy is affected by rotation. We see that  ${R}(r,\theta)\rightarrow R$ for $r>>a$. 
\begin{figure*}[t]
	\begin{centering}
		\begin{tabular}{cc }
		    \includegraphics[scale=0.81]{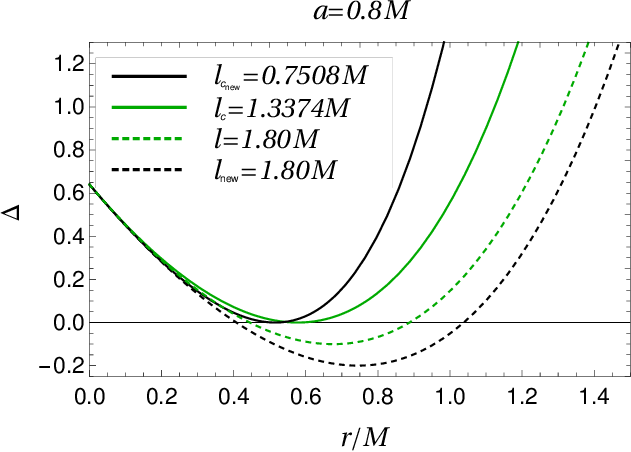}
		     \includegraphics[scale=0.81]{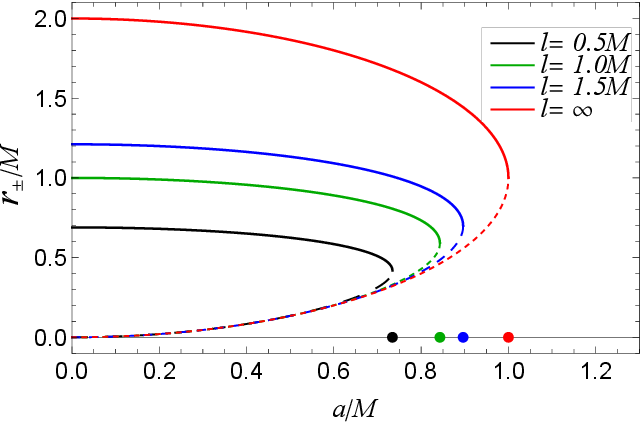}
		     \end{tabular}
	\end{centering}
\caption{Plot showing the variation of $\Delta$  with $r/M$ for both Kerr-AdS and new Kerr-AdS. The extremal black hole for Kerr-AdS occurs for  $l=1.3374M$ is at $r_+=0.62M$ whereas for new Kerr-AdS for $l=0.75084M$ is at $r_+=0.55M$  with $a=0.8$ (\textit{Left}). The event horizon (solid lines) and Cauchy horizon (dotted lines) of the new Kerr-AdS black hole are compared to the Kerr black hole ($l=\infty$), revealing their distinct horizon structures. The two horizons coincide at the points where two lines meet - extremal black holes (shown by coloured dots on the horizontal axis) (\textit{Right}).}\label{plot1}		
\end{figure*}

\begin{figure*}[t]
	\begin{centering}
		\begin{tabular}{cc}
		     \includegraphics[scale=0.8]{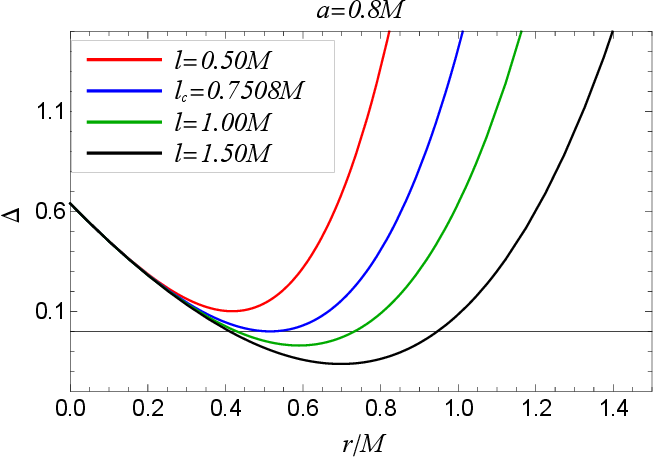}
		     \includegraphics[scale=0.8]{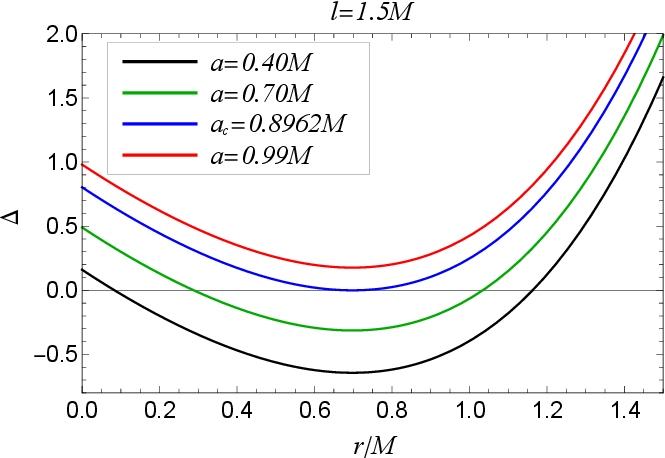}
		      	\end{tabular}
	\end{centering}
\caption{Plot showing the variation of $\Delta$  with $r/M$ for the new Kerr-AdS black holes at $a=0.8M$ and different values of $l$. The case  $l=l_c$  corresponds to an extremely new Kerr-AdS black hole (\textit{Left}). Plot showing the variation of $\Delta$  with $r/M$ for the new Kerr-AdS black holes at $l=1.5M$ and different values of $a$. The case $a=a_c$ corresponds to an extremal new Kerr-AdS black hole (\textit{Right}).}\label{plot1b}		
\end{figure*}
\begin{figure*}[t]
	\begin{centering}
		\begin{tabular}{cc}
		     \includegraphics[scale=0.8]{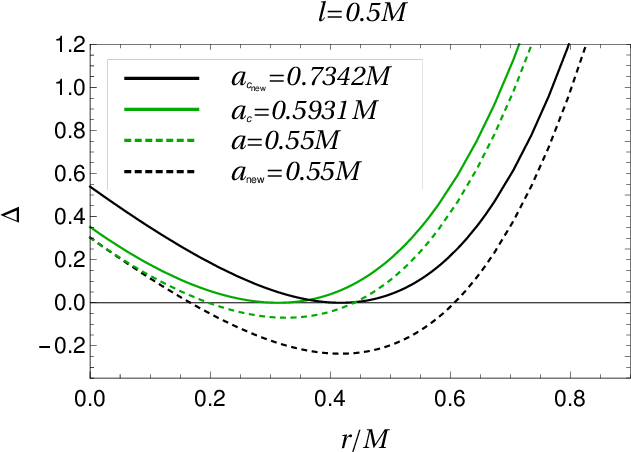}
		     \includegraphics[scale=0.8]{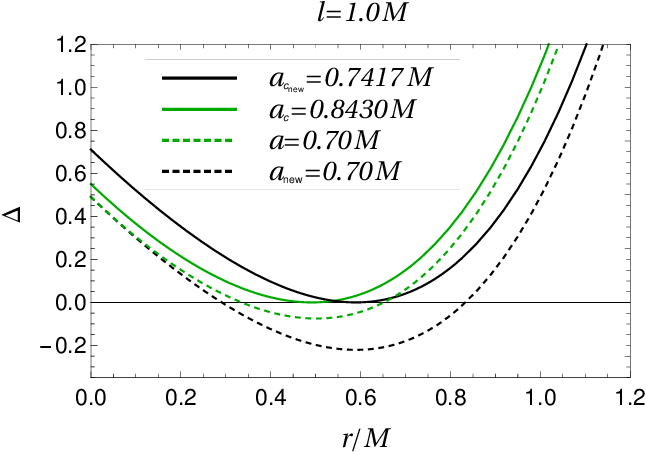}
		      	\end{tabular}
	\end{centering}
\caption{Plot showing the variation of $\Delta$  with $r/M$ for the Kerr-Ads and the new Kerr-AdS black holes at the value of $l=0.5M$ (\textit{Left}) and $l=1.0M$ (\textit{Right}). The case  $a=a_c$  corresponds to an extremel new Kerr-AdS black hole.  (\textit{Right}).}\label{plotl}		
\end{figure*}
Next, we focus on the horizon characteristics of the rotating new Kerr-AdS black holes (\ref{newKerrds}). The horizons can be determined by   the  zeroes of $g^{rr}=\Delta=0$, i.e.,
\begin{equation} \label{horizoneq}
\frac{r^4}{l^2 } + ( r^2-2 M r +a^2)= 0.
\end{equation}
where different values of parameters such as $a$ and $l$ admit two positive roots. The different allowed values of parameters,  $a$ and $l$, are illustrated in the parameter space as given in Fig~/\ref{plot1a}. The two real roots of Eq.~(\ref{horizoneq}) correspond to the innermost Cauchy horizon $r_-$ and the outer horizon $r_+$, satisfying the relationship $r_- < r_+$.
The roots of the Eq.~(\ref{horizoneq}) can be expressed as 
\begin{equation}\label{echorizon}
    r_{\pm} = \frac{1}{2}\left[ \sqrt{\frac{-2l^2}{3}+\alpha} \pm \sqrt{\frac{-4l^2}{3} + \frac{4\sqrt{3}l^2M}{\sqrt{-2l^2+3\alpha}}-\alpha}\right]
\end{equation}
with 
\begin{eqnarray}
 \alpha &=& \frac{1}{3}\left[ \left(\frac{2}{\beta}\right)^{1/3} (12 a^2 + l^2)l^2 + \left(\frac{\beta}{2} \right)^{1/3}\right] \nonumber ~~~~~~
 \beta = 2l^3\sqrt{ (\gamma l)^2 -  \left(12 a^2 + l^2\right)^3} + 2 \gamma \nonumber \\
 \gamma &=& -36 a^2+l^2+54 M^2
\end{eqnarray}
Eq.~(\ref{horizoneq}) admits two real positive roots ($r_{\pm}$) for nonzero values of $a$ and $l$. These roots are given by Eq.~(\ref{echorizon}) and correspond to Cauchy $(r_{-})$ and event horizons $(r_{+})$, which are shown in Figures ~\ref{plot1} and ~\ref{plot1b}. For the horizon to exist, the parameter $l$ needs to be larger than the critical value $l_c$ (see Figures ~\ref{plot1} and ~\ref{plot1b}). The $l_c$ results when $\Delta=0$ has two degenerate roots. For Kerr-AdS black holes, the event horizon radius is less than that of Kerr black holes (see Figures~\ref{plot1} and ~\ref{plot1b}).  As $l$ grows, the horizon radii increase, reaching a maximum value for $l=\infty$ corresponding to the Kerr horizon.  As can be seen in Figures~\ref{plot1} and ~\ref{plotl}, the Kerr-AdS black holes differ from the Kerr black hole as the former has a smaller extreme rotation parameter $a$, thus altering the strong field region's spacetime structure.

The horizon of the Kerr-AdS black hole has an extremal configuration corresponding to the minimal mass. This configuration can be found by solving the degenerate case $D=0$, where $D$ is the discriminant of the Eq.~(\ref{horizoneq}), as stated: 
\begin{equation}
D = 16 l^6 \left(a^2 \left(36 l^2 M^2+l^4\right)-8 a^4 l^2+16 a^6-l^2 M^2 \left(l^2+27 M^2\right)\right).
\end{equation}
There exists a critical of value for $M=M(l,a)$ such that $D=0$ as given by
\begin{equation}\label{bounds}
	M(a,l) =\frac{\sqrt{(36 a^2-l^2)l + \sqrt{\left(12 a^2+l^2\right)^3} }}{3 \sqrt{6l}}
\end{equation}
The  Eq.~\eqref{echorizon} possesses two distinct real solutions $l>0$ and $D<0$. Furthermore, the minimum mass of the black hole can be determined by finding the critical radius at which the derivative of the black hole mass with respect to r, denoted as $M'(r_{E},a,l)$, equals zero, i.e.,
\begin{equation}
 r_{E}=\sqrt{\frac{\alpha-l^2}{6}}. 
\end{equation}
where ${\alpha}=\sqrt{l^2 \left(12 a^2+l^2\right)}$ and for the physical solution, we take only the largest root $r_{E}$ of the extremal Kerr-AdS black hole corresponding to the minimum mass configuration. 

The metric \eqref{newKerrds}, akin to the Kerr solution, exhibits independence with respect to $t$ and $\phi$, thus possessing two Killing vectors: $\eta^\mu = \partial_t$ and $\xi^\mu = \partial_\phi$. These vectors represent time translational symmetry along the $t$ axis and rotational symmetry around the $\phi$ axis. Hence, the Killing field $\chi^\mu$ linked to these Killing vectors can be represented as $\chi^\mu = \eta^\mu + \Omega \xi^\mu$, with $\Omega$ denoting the angular velocity of the metric. The angular velocity is chosen in such a way that $\xi^\mu$ becomes null near the event horizon of the black hole, which guarantees that $\chi^\mu \chi_\mu$ equals zero \cite{Modesto:2010rv} such that
\begin{equation}
    g_{tt}+{2}{\Omega}g_{t{\phi}}+{\Omega^2{g_{{\phi}{\phi}}}}=0,
\end{equation}
where
\begin{equation}
{\Omega}=-\frac{g_{t{\phi}}}{g_{{\phi}{\phi}}}\pm\sqrt{\left(\frac{g_{t{\phi}}}{g_{{\phi}{\phi}}}\right)^2-\frac{g_{tt}} {g_{{\phi}{\phi}}}}=-{\omega}\pm\sqrt{{\omega^2}-\frac{g_{tt}} {g_{{\phi}{\phi}}}},
\end{equation} 
\begin{equation}
{\Omega}=\frac{\pm\sqrt{(a^2(a^4 + (2r^2 - 2\Delta)a^2 + r^4 - 2r^2\Delta + \Delta^2 - \Sigma)\sin^2\theta + \Delta\Sigma}-  a(a^2 +r^2 -\Delta)\sin{\theta}}{\Sigma\sin\theta}, 
\end{equation}
\begin{equation}
 {\omega}=-\frac{g_{t{\phi}}}{g_{{\phi}{\phi}}}= \frac{a(a^2 + r^2 - \Delta)}{\Sigma},
\end{equation}
at the event horizon ${\Delta({r_+})}=0$ angular velocity reduces to 
\begin{equation}
 {\Omega_{+}}={\omega}|_{r=r_{+}}=-\frac{g_{t{\phi}}}{g_{{\phi}{\phi}}} =-\frac{a}{a^2+r^2_{+}}. 
\end{equation}
The mass and angular momentum of a black hole is connected to the conserved quantities that are linked to the asymptotically timelike and spacelike Killing vector fields, denoted as $ \eta^\mu $ and $ \xi^\mu $, respectively. The quantities are defined on a spacelike hypersurface denoted as $ \varsigma_t $, a surface with a constant value of $ t $ and a unit normal vector $ n_\nu $. The hypersurface in question spans from the event horizon to spatial infinity, conforming to the definitions of conserved quantities as provided by Komar \cite{Komar:1959ab}.
 Consequently, the effective mass read as 
\begin{equation}
 M_{\rm {eff}} =\frac{1}{8\pi}\int_{S_{t}}\nabla^{\mu}\eta^{\nu}_{(t)}dS_{{\mu}{\nu}},\label{meff}
\end{equation}
where $S_t$ is a constant-$t$ and constant-$r$ surface with unit outward normal vector $\sigma_{\mu}$, and the two-boundary of the hypersurface $\Sigma_t$. $dS_{\mu\nu}=-2n_{[\mu}\sigma_{\nu]}\sqrt{h}d^2\theta$ is the surface element of $S_t$, $h$ is the determinant of ($2\times 2$) metric on $S_t$ are respectively, timelike and spacelike unit outward normal vectors.
\begin{equation}
	n_{\mu}=-\frac{\delta^{t}_{\mu}}{|g^{tt}|^{1/2}},\qquad \sigma_{\mu}=\frac{\delta^{r}_{\mu}}{|g^{rr}|^{1/2}},
\end{equation}
Hence, the mass integral~(\ref{meff}) turns into an integral over a closed 2-surface that can be stretched to the boundary of space as
\begin{align}
	M_{\text{eff}}
	=& \frac{1}{4\pi}\int_{0}^{2\phi}\int_{0}^{\phi}\frac{\sqrt{g_{\theta\theta}g_{\phi\phi}}}{|g^{tt}g^{rr}|^{1/2}}\left(g^{tt}\Gamma^{r}_{tt}+g^{t\phi}\Gamma^{r}_{t\phi} \right)d\theta d\phi,
\end{align}
The effective mass of the new Kerr-AdS black hole inside a 2-sphere with radius $r$ can be calculated using the metric (\ref{newKerrds}) given as 
\begin{equation}
 M_{\rm{eff}}=\frac{3r^2}{l^2}\left({a}+
 \frac{r^2}{a}\right)\arctan\left(\frac{a}{r} \right)-\frac{{r}^{3}}{{l}^{2}}+M.  
\end{equation}
This is corrected due to the AdS parameter $ l $ and in the absence of AdS spacetime (\( l = \infty $), it reduces to the mass of the Kerr black hole, denoted as $ M_{\text{eff}} = M $.

Next, we compute the effective angular momentum using the spacelike Killing vector $\eta^{\mu}_{(\phi)}$\cite{Komar:1958wp,Kumar:2017qws}
\begin{equation}\label{ang}
	J_{\rm{eff}}=\frac{1}{16\pi}\int_{S_t}\nabla^{\mu}\eta^{\nu}_{(\phi)}dS_{\mu\nu}.
\end{equation}
Using the definitions of surface area element, Eq.~(\ref{ang}) is expressed as
\begin{align}
	J_{\text{eff}}=&-\frac{1}{8\pi}\int_{0}^{2\phi}\int_{0}^{\phi}\nabla^{\mu}\eta^{\nu}_{(t)}n_{\mu}\sigma_{\nu}\sqrt{h}d\theta d\phi,\nonumber\\
	=& \frac{1}{8\pi}\int_{0}^{2\phi}\int_{0}^{\phi}\frac{\sqrt{g_{\theta\theta}g_{\phi\phi}}}{|g^{tt}g^{rr}|^{1/2}}\left(g^{tt}\Gamma^{r}_{t\phi}+g^{t\phi}\Gamma^{r}_{\phi\phi} \right)d\theta d\phi.
\end{align}
Upon using the new Kerr-AdS black hole metric (\ref{newKerrds}) and integrating the effective angular momentum, it becomes 
\begin{equation}\label{Jeff1}
J_{\rm{eff}}=\frac{3r^2}{l^2}\left({{r}^{2}}+\frac {a^2}{2}+\frac{r^4}{2a^2}\right)\arctan\left(
\frac{a}{r}\right)-{\frac {5a{r}^{3}}{2\,{l}^{2}}}-{\frac{3{r}^{5}}{2{l}^{2}a}}+aM.  
\end{equation}
In the  limit $l\to \infty$, the effective angular momentum Eq. (\ref{Jeff1}) restore the Kerr black hole value $J_{\rm{eff}} = M a$.

\section{ P-V criticality of new Kerr-AdS black hole}\label{sec3}
This section explores the P-V criticality of the new Kerr-AdS black holes. These aspects play a crucial role in understanding the behaviour and characteristics of these black holes within the context of thermodynamics and critical phenomena. With the introduction of quantum field theory in curved spacetime, black hole thermodynamics emerged, drawing parallels between thermodynamic quantities like internal energy, temperature, and entropy and black hole properties like mass, surface gravity, and event horizon area  \cite{Bekenstein:1973mi,Bardeen:1973gs,Bekenstein:1972tm,Bekenstein:1974ax}. The terminology used to describe black hole phenomena has evolved in modern scientific discourse,
 particularly in the context of extended Anti-de Sitter  spacetime, where it is referred to as black hole chemistry \cite{Kubiznak:2014zwa,Kubiznak:2016qmn}. This new perspective treats the cosmological constant $\Lambda$ as a dynamic thermodynamic pressure \cite{Teitelboim:1985dp,Brown:1987dd}, denoted by $P=3/8 \pi l^2$. Considering $\Lambda$ as a thermodynamic pressure ($P$) leads to reinterpreting black hole mass as enthalpy ($H=m$) instead of internal energy \cite{Kastor:2009wy, Dolan:2011xt,Dolan:2010ha}.

Consequently, in the case of a new Kerr-AdS black hole, the corresponding black hole thermodynamic volume has the following form:
\begin{equation}
 V=\left(\frac{\partial M}{\partial P}\right)_{S,J}\,=\frac{4}{3}\pi r_+(r_+^2+a^2),
\end{equation}
with $r_+$ being the event horizon of the black hole. The thermodynamic volume for various known black hole solutions was studied in \cite{Cvetic:2010jb}. We may compute other thermodynamic information when the pressure and volume have been determined. We may then investigate adiabatic comprehension, specific heat at constant pressure, and specific heat at constant volume \cite{Dolan:2010ha, Dolan:2011xt, Dolan:2011jm}. Additionally, as evaluated by Dolan \cite{Dolan:2011xt}, this presents an interesting opportunity to reevaluate the AdS black hole's critical behaviour in an extended phase space that considers pressure and volume as thermodynamic factors.  Specifically, Dolan \cite{Dolan:2011xt} examined the rotating charged AdS black hole's $P=P(V,T)$ equation of state, found an analogy with the vdW  $P-V$ diagram, and identified its critical point. By identifying with the liquid-gas system in the extended phase space, we explore the $P-V$ critically of new Kerr-AdS black holes to shed light on the system's previously reported vdW -like behaviour, as detailed in \cite{Chamblin:1999tk, Chamblin:1999hg}. 
\begin{figure*}[t]
	\begin{centering}
		\begin{tabular}{p{9cm} p{9cm}}
		    \includegraphics[scale=0.75]{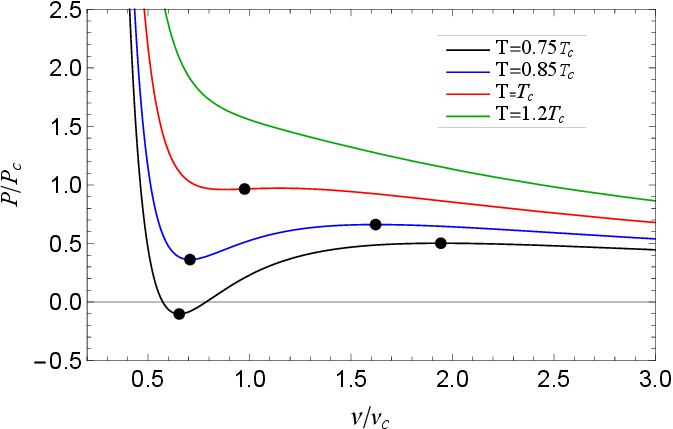}&
		     \includegraphics[scale=0.75]{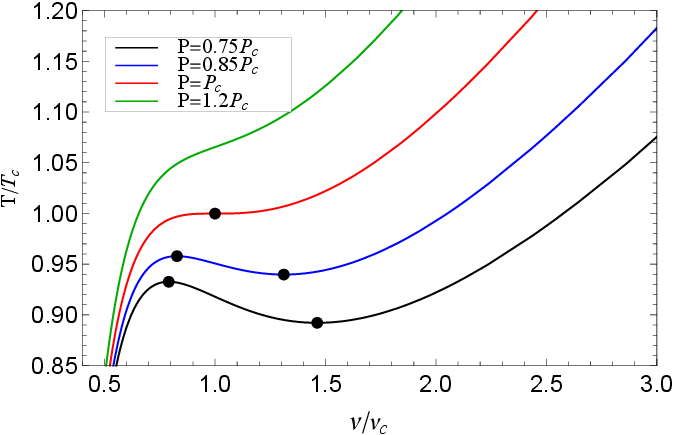}\\
		     \end{tabular}
	\end{centering}
	\caption{Plot showing the variation of reduced pressure with reduced specific volume (isotherm). The black dots indicate the value at which $(\partial_{v}P)_{T} =0$ (\textit{Left}). Plot showing the variation of reduced temperature with reduced specific volume (isobar). The black dots indicate the value at which $(\partial_{v}T)_{P} =0$  (\textit{Right}).} \label{fig3}
\end{figure*}
Because of the nonzero size of fluid molecules and their attraction to one another, the vdW equation, e.g. \cite{Goldenfeld:1992qy}, is a good approximation of the ideal gas law, which approximates the behaviour of actual fluids. The basic qualitative characteristics of the liquid-to-gas phase transition are frequently described using it. The  vdW  equation has the following form
\begin{equation}
\left(P+\frac{a}{v^2}\right)(v-b)=kT\label{van},
\end{equation}
where $k$ is the Boltzmann constant, $T$ is the temperature, $v=V/N$ is the specific volume, and $P$ is the fluid's pressure.  To express the attractive force between molecules and the fluid's limited molecular size, $a$ and $b$ are constants. Subsequently,  we can also write the equation of the state of the new Kerr-AdS black hole in the form of $P=P(V,T)$  as
\begin{equation}\label{state}
P= \frac{T_+}{2 r_+}\left(1+\frac{a^2}{r_+^2}\right)+\frac{a^2}{8 \pi  r_+^4}-\frac{1}{8 \pi r_+^2},
\end{equation}
Here $P$ is the pressure, $V$ is the thermodynamic volume, $T$ is the temperature of the black hole at the event horizon $r_+$, and $a$ is its rotation parameter. Understanding the behaviour of the isothermal curve in the $P-V$ plane is often necessary to investigate the VdW-like phase transition. Nevertheless, we see that the pressure $P$ is difficult to be expressed in terms of $T$, $V$, and $a$ based on Eq. (\ref{state}). We, however, after comparing with the vdW  Eq. \eqref{van}, specific volume $v$ can be interpreted as $v=2r_+$ and re-expressing  Eq. \eqref{state} in terms of $v$, we have 
\begin{equation}\label{EoS}
 P= \frac{T}{v}\left(1+\frac{4 a^2}{v^2}\right)+\frac{2 a^2}{\pi v^4}-\frac{1}{2 \pi v^2}.
\end{equation}
From the $P-V$ diagram in Fig.~\ref{fig3}, there would be an inflexion point at $T_c$, called {\em critical point}, similar to the vdW fluid, below which the pressure $P$ becomes negative for some horizon radius  $r_+$. According to Maxwell's area law, this behaviour is nonphysical or unrealistic in the fluid, as one has to replace the oscillatory part of the isotherm with an isobar.  The analytical solution of the above conditions leads to critical points. In thermodynamic systems, variables such as pressure and temperature can exhibit complex behavior that leads to multiple extremal points—specific locations where the system's properties reach maximum or minimum values (cf.~Fig.~\ref{fig3}). These extremal points,  depicted as distinct black dots in Fig.~\ref{fig3}, are crucial in understanding the nuanced changes within the system. These  points are critical in understanding phase transitions, as they indicate the conditions under which a system undergoes a significant change in its state. A first-order phase transition is typically marked by two extremal points, representing a discontinuous jump in a thermodynamic quantity. This duality signifies the coexistence of two phases with distinct properties at the transition temperature or pressure (cf.~Fig.~\ref{fig3}). In contrast, a second-order phase transition, which is continuous and is characterized by a single extremal point. However, it is crucial to recognize that the absence of a single extremal point does not necessarily preclude the occurrence of a phase transition. We obtained the critical points by using the condition $(\partial_{S}T)_{P,a} = (\partial_{S,S}T)_{P,a}=0$.   Solving them, we were able to determine the analytical critical points for the new Kerr-AdS black hole as follows:
\begin{equation}\label{criticalvaluesAdS}
 T_c=\frac{0.0309833}{a}\,,\quad P_c=\frac{0.00180781}{a^2}\,,\quad r_c=3.10217a\,.
\end{equation}
The critical radius given above corresponds to the critical thermodynamic volume $V_c$, Gibbs energy $G_c$, and critical entropy $S_c$ as
\begin{equation}\label{criticalvaluesAdS2}
V_c=138.045 a^3\,,\quad G_c=0.904275a\,,\quad S_c=33.3746a^2\,.
\end{equation}
For the new Kerr-AdS black hole, the universal constant $\epsilon$ was determined to be 
\begin{equation}
\epsilon = \frac{P_c V_c^{1/3}}{T_c}=0.301555\,,
\end{equation}
which is a bit less than 3/8–the value for vdW fluid.
\begin{figure*}[h]
	\begin{centering}
		\begin{tabular}{p{9cm} p{9cm}}
		    \includegraphics[scale=0.75]{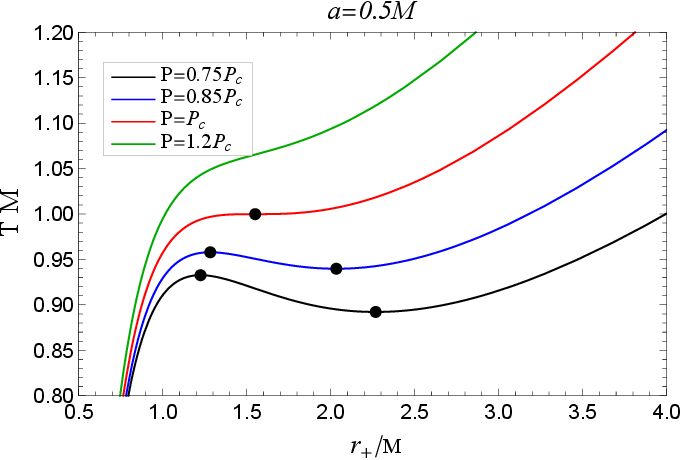}&
		     \includegraphics[scale=0.75]{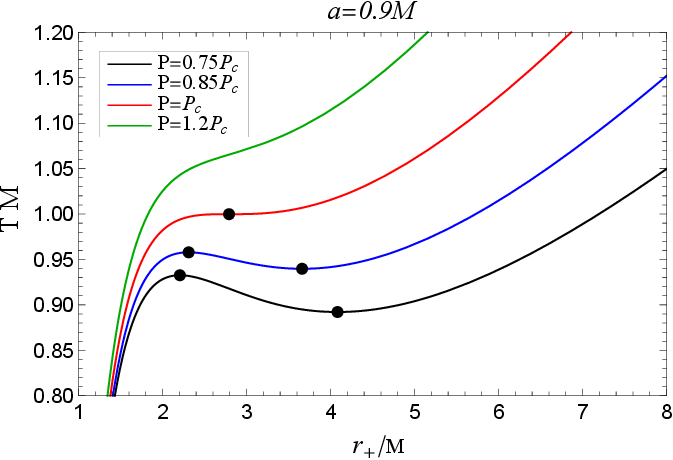}\\
		     \end{tabular}
	\end{centering}
	\caption{Plot showing the variation of reduced temperature with the reduced event horizon of the new Kerr-AdS black hole is at $a=0.5$ (\textit{Left}). The plot showing the variation of reduced temperature with the reduced event horizon of the new Kerr-AdS black hole is at $a=0.9$ (\textit{Right}). In both plots, the black dots indicate the value at which $(\partial_{r_+}T)_{P} =0$.} 
	\label{fig7}
\end{figure*}
\begin{figure*}[h]
	\begin{centering}
		\begin{tabular}{p{9cm} p{9cm}}
		    \includegraphics[scale=0.75]{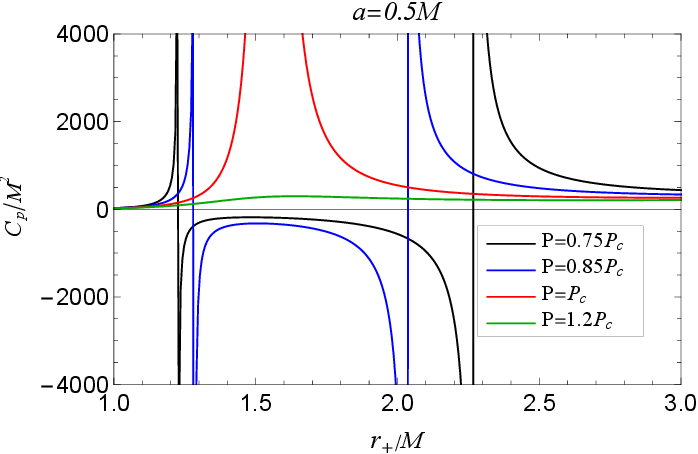}&
			\includegraphics[scale=0.75]{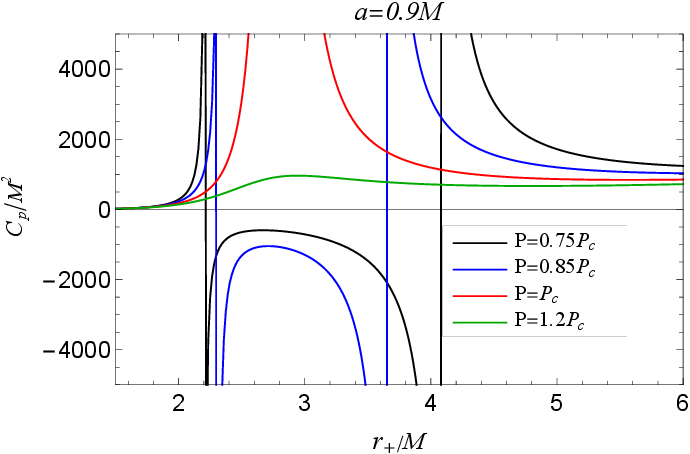}\\
			\end{tabular}
	\end{centering}
	\caption{Plot showing the variation of $C_p/M^2$ with horizon radius $r_+/M$ is at $a=0.5$ (\textit{Left}). Plot showing the variation of $C_p$ with horizon radius is at $a=0.9$ (\textit{Right}).} \label{fig5}		
\end{figure*}
\section{ Horizon thermodynamics and Stability } 
The thermodynamic properties of black holes are closely related to their event horizons. For new Kerr-AdS black holes, horizon thermodynamics involves studying quantities such as temperature, entropy, heat capacity and Gibbs free energy at the event horizon $(r_{+})$.
Now, we will investigate the basic thermodynamic properties of the new Kerr-AdS black hole. The black holes metric \eqref{newKerrds}  can be characterized by the  Arnowitt-Desser-Misner (ADM) mass \cite{Arnowitt:1962hi} $(M_{+})$, which can be expressed in terms of the event horizon radius $(r_{+})$ by solving  $g{^{rr}}{(r_+)}=\Delta{(r_+)}=0$ as
\begin{equation}
M_{+}=\frac{1}{2r_+}\left(a^2 + r^2_{+}+\frac{r^4_+}{l^2}\right).
\end{equation}
In the limit $l \to \infty$, the above expression reduces to $(a^2 + r^2_{+})/2r_+$, in the context of the Kerr black hole limit, black holes behave as thermodynamic entities. A black hole's temperature $ T $ can be determined from the surface gravity $ \kappa $ evaluated at the Killing horizon \cite{Fodor:1996rf}. Hawking demonstrated that the surface gravity $ \kappa $ is directly related to the black hole's temperature at the event horizon.

\begin{eqnarray}\label{temp}
T_{+} = {{\kappa}\over{2\pi}}= \frac{{\Delta'{(r_{+})}}}{{4\pi}(a^2+r^2_+)}  \implies   \frac{r_{\rm +}}{4\pi(r_{\rm +}^{2}+a^{2})} \left(1-\frac{a^{2}}{r_{\rm +}^{2}}+3\frac{r_{\rm +}^{2}}{l^{2}}\right).
\end{eqnarray}
where $'$ denotes the differentiation with respect to $r$. Fig. \ref{fig7} displays the temperature graph as a function of horizon radius, with varied values of parameter $P$. When $P < P_C$, there is a non-monotonic pattern with one local maximum and one minimum. The two extreme points get closer and meet at $P = P_C$, resulting in a reflection point corresponding to the critical point. Further, we can see that the temperature of the Kerr black hole 
\begin{equation}
T^{K}_{+} = \frac{(r_{\rm +}^{2}-a^{2})}{4\pi r_{\rm +} (r_{\rm +}^{2}+a^{2})}.    
\end{equation}
can be recovered in the $l\to\infty$. We now proceed to find the entropy of the black hole from Bekenstien's area law \cite{Jacob:2333abc}, the entropy of the black hole is multiple of the surface area of the event horizon as $S_+={A_+/4}$. 
\begin{equation}
 S_+=\frac{1}{4}\int_{0}^{2\pi}\int_{0}^{\pi}\sqrt{g_{\phi\phi}g_{\theta\theta}}{d\theta}{d\phi}|_{r=r_{+}}=\pi(r^2_{+}+a^2),
\end{equation}

Next, we examine the phase transition of the new Kerr-AdS black holes. Evaluating the heat capacity at constant pressure, denoted as $C_P$ is pivotal in assessing the stability of a thermodynamic system. A positive heat capacity ($C_P>0$) signifies the thermodynamic stability of the black hole, whereas a negative heat capacity ($C_P<0$) indicates thermodynamic instability within the system. The Divergence of heat capacity indicates phase transitions of a black hole. By definition specific heat at constant pressure $C_{p}={T}\left(\frac{\partial{S}}{\partial{T}}\right)_{p}$ 

\begin{equation}
C_{P}=\frac{\pi  \left(a^2+r_+^2\right) \left(a^6+3 a^4 r_+^2+16 \pi  P r_+^8 \left(5-4 \pi  a^2 P\right)+r_+^6 \left(5-32 \pi  a^2 P\right)+a^2 r_+^4 \left(16 \pi  a^2 P-9\right)+192 \pi ^2 P^2 r_+^{10}\right)}{2 r_+^2 \left(4 \pi  P r_+^2+1\right) \left(a^4+r_+^4 \left(24 \pi  a^2 P-1\right)+4 a^2 r_+^2+8 \pi  P r_+^6\right)}.
\end{equation}
A black hole is stable if its heat capacity is positive ($C_P > 0$), and unstable if the heat capacity is negative ($C_P < 0$). Fig. \ref{fig5} depicts the heat capacity for different values of $P $. We observe two distinct behaviours: when the heat capacity is positive ($P > P_C$), the black hole is thermodynamically stable. Conversely, when the heat capacity is positive, negative, or divergent for $P \leq P_C$, the black hole is experiencing a phase transition \cite{Hawking:1982dh,Davies:1977bgr}. 

To further analyze the behaviour, we plot the heat capacity $\tilde{C_{P}} $ against $r_{+} $ in Fig.~\ref{fig5} for various values of $P $ and $a $. It is observed that there exists a pressure $P $, given a specific $a $, at which the heat capacity shows discontinuity at critical radii $r_{c1} $ and $r_{c2} $, corresponding to local maxima and minima of the Hawking temperature, with $r_{c1} < r_{c2} $ when $P < P_c $ (Fig.~\ref{fig5}). This discontinuity indicates a second-order phase transition \cite{Hawking:1982dh,Altamirano:2014tva}. As depicted in Fig.~\ref{fig5}, three different cases are evident:

1. When $P < P_c $, a second-order phase transition occurs between small black holes (SBH) and intermediate black holes (IBH), and between IBH and large black holes (LBH). Here, black holes with $r_+ < r_{c1} $ (SBH) and $r_+ > r_{c2} $ (LBH) are locally stable with positive specific heat, while those with $r_{c1} < r_+ < r_{c2} $ and negative specific heat are locally unstable.
   
2. For $P > P_c $, no phase transition occurs, and the specific heat remains positive, indicating the local stability of the black holes.
   
3. The case of $P = P_c $ corresponds to the coexistence of two stable black holes at the inflexion point $r_+ \approx 1.5524M $ for $a = 0.5M $.

\begin{figure*}[t]
	\begin{centering}
		    \includegraphics[scale=0.85]{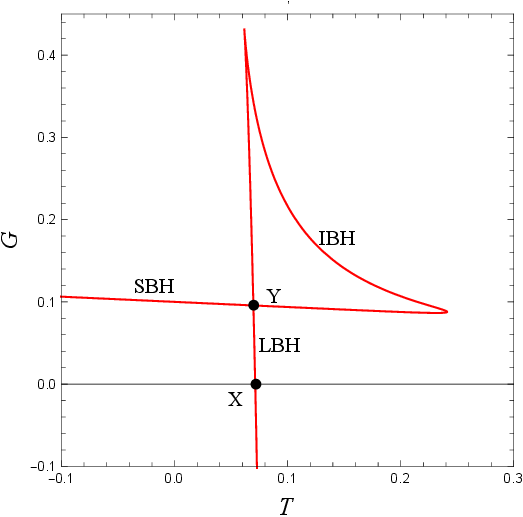}
	\end{centering}
\caption{Swallowtail structures in $G-T$ plots for AdS black holes, illustrating phase transitions and thermodynamic stability. Small, intermediate, and large black holes are denoted by the letters SBH, IBH, and LBH, respectively. }\label{fig7a}		
\end{figure*}
Interestingly, when $a=0.9M$, at $P=0.85 P_C$, the discontinuity in the heat capacity occurs at $r_{c1}=1.2839M$ and $r_{c2}=2.0342M$, where not only does the Hawking temperature reach its maximum (minimum) value $T_{\text{max}}=0.95785/M$ ($T_{\text{min}}=0.93975/M$), but the heat capacity also diverges. It indicates a phase transition from lower to higher mass, corresponding to the black hole's change from positive to negative heat capacity. The black hole is unstable between $r_{c1}$ and $r_{c2}$.

Investigating the thermodynamic stability of black holes entails examining the behaviour of Gibbs free energy \cite{Kumar:2020cve, Tzikas:2018cvs}. We focus on the regions with negative free energy, indicating when black holes are thermally favoured compared to the reference background. The calculation of the black hole's free energy is derived from \cite{Kumar:2020cve, Tzikas:2018cvs, Altamirano:2014tva}:
Further, the Gibbs free energy $G_+=H-T_+S_+$, which is globally minimized by the equilibrium state of the system, elucidates the phase structure of the black hole and its stability. Now, we will study the phase structure of the new Kerr black hole. It is obtained as
\begin{equation}
 G_{\rm +}=\frac{(3 a^2+r^2_{+})l^2-r^4_{+}}{4 r_{+} l^2}.    
\end{equation}

\begin{figure*}[t]
	\begin{centering}
		\begin{tabular}{cc}
		    \includegraphics[scale=0.8]{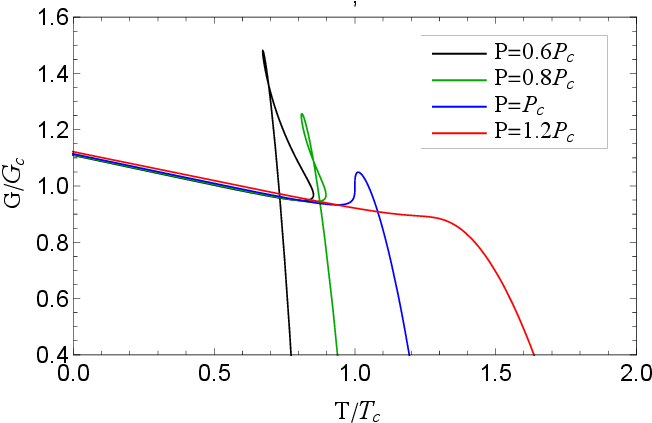}
		    \includegraphics[scale=0.8]{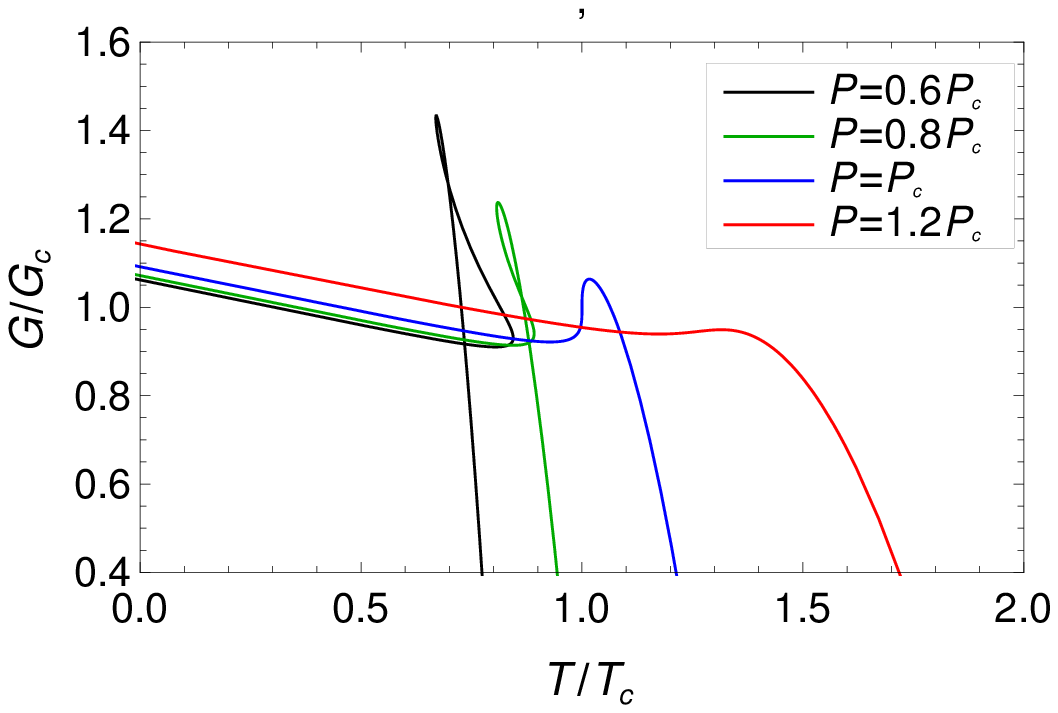}
			\end{tabular}
	\end{centering}
	\caption{Plot showing the variation of reduced Gibbs free energy with reduced temperature for the new Kerr-AdS black hole (\textit{Left}) and for Kerr-AdS black holes (\textit{Right}). Both exhibit swallowtail behavior due to phase transitions} \label{gt}		
\end{figure*}
This equation enables us to examine the circumstances in which black holes exhibit thermodynamic stability. The system's global stability is evaluated using Gibbs free energy, where the global minimum represents the desired state of the black hole \cite{Kubiznak:2016qmn}. The Gibbs free energy is crucial for determining first-order phase transitions by analyzing the swallowtail behaviour found in the $\tilde{G}$-$\tilde{T}$ plots. The curve in Figures \ref{fig7a} and \ref{gt} demonstrates the correlation between Gibbs free energy and temperature. When the pressure $\tilde{P}$ is lower than the critical pressure $\tilde{P}_c$, the $\tilde{G}-\tilde{T}$ diagram exhibits a swallowtail structure \cite{Nam:2019clw}, where the curve representing the Gibbs free energy intersects itself. This characteristic signifies a phase transition of the first order between  SBH and LBH, as illustrated in Figures \ref{fig7a} and \ref{gt}. Examining the Gibbs free energy demonstrates that when the pressure is below the crucial threshold ($\tilde{P}<\tilde{P}_c$), the graph displays the distinctive swallowtail pattern, indicating a first-order phase transition between SBH and LBH. More precisely, when $\tilde{P}<\tilde{P}_c$, the $G$ -- $T$ graph displays three distinct states of black holes: stable black hole, unstable intermediate black hole, and stable large black hole (see Figures \ref{fig7a} and \ref{gt}).

The Gibbs free energy displays a typical swallowtail behaviour (cf. Fig.~\ref{fig7a}), which in turn depicts the existence of the first-order small-large black hole phase transition. The typical swallow tail behaviour of the Gibbs free energy with fixed spin parameter  $a$ and pressure $P$ is plotted in the $G-T$ plane in Fig. \ref{fig7a}. It is evident from the figure that the Gibbs free energy disappears at point $X$. The system will initially follow the small black hole branch as temperature increases until point $Y$, where both small and large black holes coexist since they have the same Gibbs free energy. As the temperature increases, the system will prioritise the large black hole with a lower Gibbs free energy over the smaller and intermediate black holes. The shift from small to large black holes resembles the phase transition from liquid to gas in the Van der Waals fluid. To fully examine the transitory characteristics of black holes, regardless of their spin, we have opted to utilize reduced parameters that are not influenced by the rotation parameter $a$. Figure ~\ref{gt} shows the plot of the reduced Gibbs energy $G_r={G}/{G_c}$ related to the reduced temperature $T_r$ at a constant pressure $P$.  We have discovered a zeroth-order phase transition in the new Kerr-AdS solution, which is quite intriguing.

\subsection{ Maxwell equal area law  and co-existence curve}
\begin{figure*}[t] 
	\begin{centering}
		\begin{tabular}{cc}
            \includegraphics[scale=0.98]{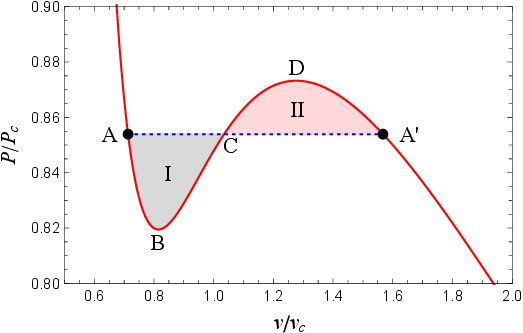}
		    \includegraphics[scale=0.98]{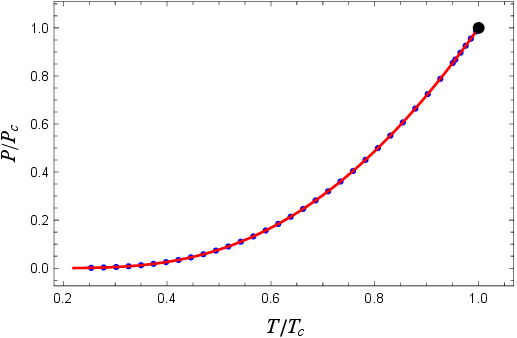}
			\end{tabular}
\end{centering}
\caption{Maxwell's Area Law for the new Kerr-AdS black hole with $T=0.95$. The phase transition point is at $P/P_c=0.85392$. The difference between the two areas I and II is $6.5794 \times 10^{-7}$ (\textit{Left}). The co-existence curve in the $\tilde{P}-\tilde{T}$ plane. The discrete points denote the numerical values and the solid line is obtained from the fitting formula Eq.~\ref{coexisC} (\textit{Right}).} \label{fig8}
\end{figure*}

The phase transition point is typically identified through the swallowtail behaviour of the Gibbs free energy. Alternatively, it can also be determined using the Maxwell equal area law.
The Maxwell equal area law for Kerr-AdS black holes is a thermodynamic law that describes the phase transition between different configurations of black holes. We have illustrated the isotherm in the $v-P$ plane in Fig. \ref{fig8} where an oscillatory behaviour exists.
Maxwell's equal area law states that during a phase transition, this oscillating part can be replaced by an isotherm in a $P-v$ diagram or isobar in a $T-S$ diagram to describe it in such a way that the areas above and below the isotherm or isobar are equal to each other for the coexisting phases. We will use the equal area law in the $P-v$ plane to investigate the phase transition in detail.  The equation of state (\ref{EoS})  for Kerr-AdS black holes relates the pressure $P$, volume $v$, and temperature $T$. Using the critical points in Eq's~(\ref{criticalvaluesAdS}) and (\ref{criticalvaluesAdS2}), we can rescale the  
the thermodynamics terms such that $p=P/ P_{C}$, $t=T/ T_C$ and $v=v/ v_C$,  the state equation can be written as 
\begin{eqnarray}
p = \frac{2.76359~t v^3+0.287172~t v-2.28807~v^2+0.23776}{v^4}
\end{eqnarray}

For $t = 0.95$ we plot the $p-v$ graph in Fig.~\ref{fig8}. An isobar replaces the oscillating component such that the areas designated by Area I and Area II—above and below the isobar—are exactly equal. This phase transition is first-order. We investigate the isobar $p^{*} = p_A=p_{A'}=0.854$; the left cross point is $v_A=0.71312$ indicates the SBH phase volume; the right cross-point $v_{A'}=1.5648$ indicates the LBH phase volume.

Usually maintained constant temperature, the Maxwell Area law construction is implemented in the $(P, V )$ plane. Since the Gibbs free energy stays constant during the phase transition, theoretically, it can be obtained from the variation in Gibbs free energy.  

\begin{eqnarray}
dG= -SdT+vdP .
\end{eqnarray}
Now, we integrate it to
\begin{eqnarray}\label{int}
 -\int^{T_{A'}}_{T_{A}} SdT+\int^{P_{A'}}_{P_{A}} vdP =0.
\end{eqnarray}
We examine the isotherm on the $P-v$ plane. Consequently, the initial component in Eq.~(\ref{int}) disappears, and in terms of reduced quantities, we obtain
\begin{eqnarray}
 \int^{p_{A'}}_{p_A} v \, dp = 0. \label{vp0}
\end{eqnarray}
The two states are known to have the same pressure, $p_{A'} = p_{A} $, along the coexistence curve. The Eq.~(\ref{vp0}) containing the integral can then be rewritten as:
\begin{eqnarray}\label{interalVdP}
 \left( \int^{p_{C}}_{p_{A}} v \, dp \right) +
 \left( \int^{p_{A'}}_{p_{C}} v \, dp \right) = 0.
\end{eqnarray}
The phase transition pressure is the same as the pressure $p_{C} = p_{A} = p_{A'} $. Equation (\ref{interalVdP}) clearly shows that the area of region the first bracket denotes me, and the negative area of region II is represented by the second one, as can be seen clearly in Fig. \ref{fig8}. As a result, we get:
\begin{eqnarray}
 \text{Area I} = \text{Area II}.
\end{eqnarray}

Consequently, this equal area law remains unchanged during the phase transition. However, the value of the phase transition parameter can also be found by applying this area rule. Remember that the same equal area law applies to the specific volume $v$.  The co-existence curve separates the two equilibrium states of any system. Whenever two surfaces with the same G-value meet, a phase transition occurs. It is governed by the slope of $\frac{\partial P}{\partial T}$ have known as Clausius-Clapeyron equation \begin{equation}
    \frac{\partial P}{\partial T}=\frac{\Delta S}{\Delta v},
\end{equation}
where $\Delta S$ and $\Delta v$ differ in entropy and specific fluid volume in two phases. 

There are two ways to construct a co-existence curve \cite{Kubiznak:2012wp}, by extrapolating Maxwell's area law or by finding a curve in the $P-T$ plane for which the Gibbs free energy and vdW temperature coincide for two different thermodynamic volumes. We had affirmed with the latter one to create the $\tilde{P}-\tilde{T}$ (reduce temperature and pressure) plane phase diagram or co-existence curve (cf. Fig.~\ref{fig8}). As a result, we can fit it using the $\tilde{T}$ polynomial and its parametrization form is:
\begin{eqnarray}\label{coexisC}
\tilde{P} &=& 0.023111 - 0.56696 \tilde{T} + 5.7954 \tilde{T}^2 - 31.619 \tilde{T}^3 + 98.79 \tilde{T}^4 - 183.77 \tilde{T}^5 + 226.63 \tilde{T}^6 - 188.15 \tilde{T}^7 \\ \nonumber &+& 101.5 \tilde{T}^8 -32.222 \tilde{T}^9 + 4.5735 \tilde{T}^{10},  
\end{eqnarray}
Thus, the coexistence curve ends at the critical point, above which it is impossible to distinguish between the phases of black holes, and has a positive slope throughout. In Fig. \ref{fig8}, we plot the numerical values of the co-existence curve  (the discrete points) and fitting formula (the solid line). We may determine how the physical variables change when phase transitions occur using this parameterized form of the coexistence curve in Eq.~(\ref{coexisC}). 
\section{Geodesics for New Kerr-AdS black holes, Photon orbit and Phase transition} \label{sec4}
Having discussed the connection between thermodynamic phase transitions and black hole horizons, it becomes crucial to expand this exploration to include the relationship between thermodynamic phase transitions and unstable photon orbits.
Accordingly, we would investigate the geodesics of a free photon orbiting the new Kerr-AdS black hole. It is generally known that in a stationary axisymmetric spacetime, the entire geodesic line will be in the equatorial plane $\theta=\frac{\pi}{2}$ provided that the initial position and tangent vector of the line are in that plane. Moreover, coordinate transformation may convert every geodesic line into an equatorial plane motion. Therefore, without loss of generality, we consider that the whole trajectory of the photon is limited on the equatorial plane. Consequently, the  metric (\ref{newKerrds}) reduces to
\begin{eqnarray}
 ds^{2}=A(r)dt^{2}-B(r)dr^{2}-C(r)d\phi^{2} + D(r)dtd\phi,
\end{eqnarray}
with the metric functions given as
\begin{eqnarray}
 A(r)=\frac{\Delta-a^2}{r^2},\quad
 B(r) =	\frac{{r}^{2}}{{\Delta}},~~~~
 C(r)=\frac{(r^2+a^2)^2-\Delta{a}^{2}}{{r}^{2}},\quad
 D(r)=\	\frac{2\, {a}}{{r}^{2}}\left(r^2+a^2-\Delta\right),
\end{eqnarray}
and $\Delta$ as provided in Eq.~(\ref{deltaN}). Furthermore, the Lagrangian governing the motion of the particle is
\begin{equation}
2\mathcal{L}=g_{\mu\nu}\dot{x}_{\mu}\dot{x}_{\nu}= A(r)\dot{t}^{2}-B(r)\dot{r}^{2}-C(r)\dot{\phi}^{2} + D(r)\dot{t}\dot{\phi}
\end{equation}
\begin{figure*}[t]
	\begin{centering}
		\begin{tabular}{cc}
		    \includegraphics[scale=0.8]{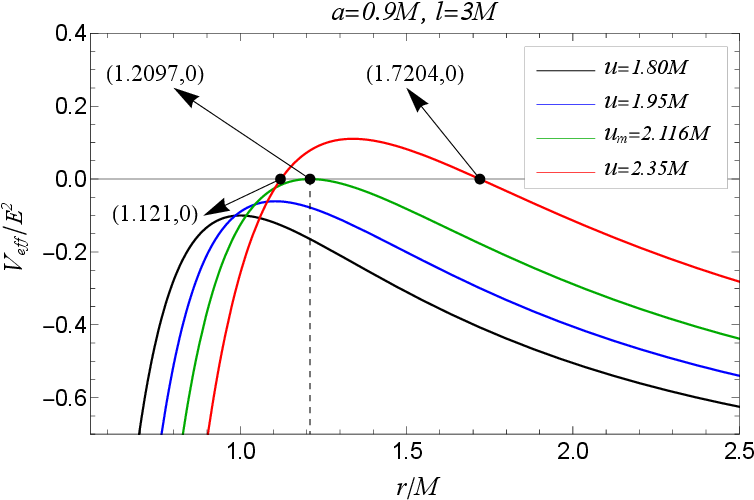}
			\end{tabular}
	\end{centering}
	\caption{A dimensionless effective potential $V_{\textit{eff}}/E^2$ of a light ray as a function of the radial coordinate $r$ at fixed values of $a$ and $l$. With $a=0.9M$ and $l=3M$, a ray of light with $u_m=2.116$ reaches the photon sphere at $r_{ps}=1.2097M$. The ray of light with an impact parameter of $u=1.11 u_m=2.35M$ will pass slightly outside the photon sphere such that it will reach a minimum distance of $r_0=1.7204 M$ with an event horizon at $r_h=1.121M$. } \label{fig6a}		
\end{figure*}
where the dot represents the derivative with respect to geodesic affine parameter. The geodesic of the new Kerr-AdS black hole spacetime in the equatorial plane has two integral constants, $E$ and $L$, connected with the Killing fields $\eta^{\mu} = \partial_t$ and $\xi^{\mu} = \partial_\phi$. The particle's energy and orbital angular momentum are $E$ and $L$. From the Lagrangian, one may obtain the canonical momentum, i.e., $p_{\mu} = g_{\mu\nu} \dot{x}^{\nu}$ of the particle as follows:
\begin{eqnarray}\label{EL}
 -E=p_{t}=-A\dot{t}-\frac{D}{2}\dot{\phi}, ~~~~~L=p_{\phi}=-\frac{D}{2}\dot{t}+C\dot{\phi}.
\end{eqnarray}
Solving them, we obtain the $t$-motion and $\phi$-motion as:
\begin{eqnarray}
\dot{t}= \frac{2 (2 E C(r)- L D(r))}{4 A(r) C(r)+D(r)^2},~~~~~
\dot{\phi}= \frac{2 (2 L A(r)+E D(r))}{4 A(r) C(r)+D(r)^2}
\end{eqnarray}
The equation of motion along the radial direction may be obtained  by applying the normalization condition $g_{\mu\nu}\dot{x}^{\mu}\dot{x}^{\nu}=-\delta^{2}$, where $\delta^{2}=0$ for null geodesics and using Eq.~(\ref{EL}) simply as follows: 
\begin{eqnarray}
 \dot{r}^{2}+V_{\textit{eff}}=0,
\end{eqnarray}
with $V_{\textit{eff}}$ being the effective potential given by 
\begin{equation}\label{Veff1}
V_{\textit{eff}}=\frac{4[L^2A(r)+E(LD(r)-EC(r))]}{B(r)[4A(r)C(r)+D(r)^2]}. 
\end{equation}
To simplify the calculations, we assign $E$=1. This is comparable to expressing  $L\rightarrow LE$,  $V_{\textit{eff}}\rightarrow V_{\textit{eff}}/E^{2}$, and introducing a new variable $u$ defined as the ratio of $L$ to $E$, which represents the impact parameter. The effective potential $V_{\textit{eff}}$ varies with the parameter $r$ and is influenced by the parameters $u$, $l$, and $M$ (or $u$, $P$, and $r_{+}$). The effective potential is displayed in Fig.~\ref{fig6a} for the new Kerr-Ads black hole. The black hole has fixed parameters with a value of $a=0.9M$ and $l=3M$ (equivalently $P=0.0132629$), but the impact parameters are allowed to vary.  By observing this figure, it is evident that each curve exhibits a peak $V^{max}_{\textit{eff}}$ at $r_{max}$, which consistently grows as $u$ increases.  Due to the constraint $\dot{r}^{2}>0$, the photon is limited to a specific range in the $r$-direction where the effective potential is such that $V_{\textit{eff}}\le 0$. If the maximum effective potential, $V^{max}_{\textit{eff}}$, is negative, an incoming photon from infinity will inevitably be pulled into the black hole. Conversely, if $V^{max}_{\textit{eff}}$ is positive, the photon will bounce back instead. The intriguing occurrences manifest at the precise moment when the peak reaches zero. In this scenario, the photon experiences a loss of both its radial velocity and acceleration. However, if it possesses a non-zero transverse velocity, it can orbit the black hole once, twice, or even several times. Therefore, in this particular scenario, when $r = r_{max}$, it is referred to as the photon orbit. From now on, we will represent it as $r = r_{ps}$. The following conditions determine this value:
\begin{eqnarray}
 V_{\textit{eff}}(r,u)\Big|_{r=r_{ps},u=u_{ps}} =  \partial_{r}V_{\textit{eff}}(r,u)\Big|_{r=r_{ps},u=u_{ps}}=0.
\end{eqnarray}
\begin{figure*}[t]
	\begin{centering}
		\begin{tabular}{cc}
	 	    \includegraphics[scale=0.8]{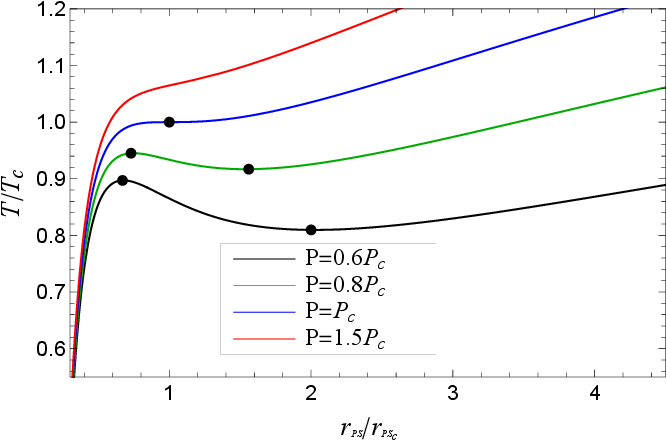}
		    \includegraphics[scale=0.8]{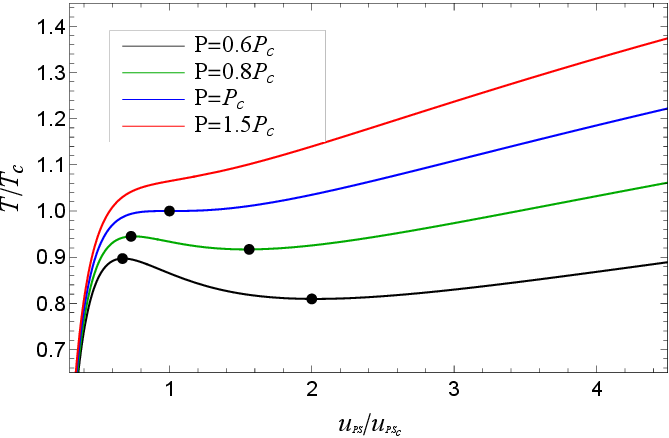}
			\end{tabular}
	\end{centering}
\caption{The behaviour of reduced temperature as a function of the radius of the photon sphere $r_{ps}/r_{psc}$ (\textit{left}) and critical impact parameter $u_{ps}/u_{psc}$ (\textit{right}) at different values of pressure. The  black dots  correspond to the extremal points in both plots. } \label{fig12}	
\end{figure*}
\begin{figure*}[t]
	\begin{centering}
		\begin{tabular}{cc}
	 	    \includegraphics[scale=0.8]{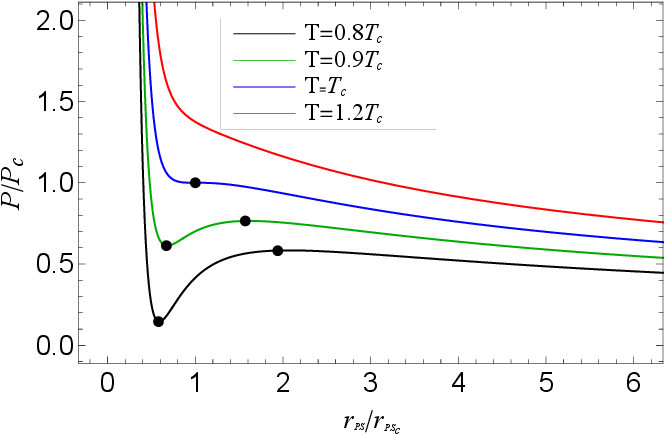}
		    \includegraphics[scale=0.8]{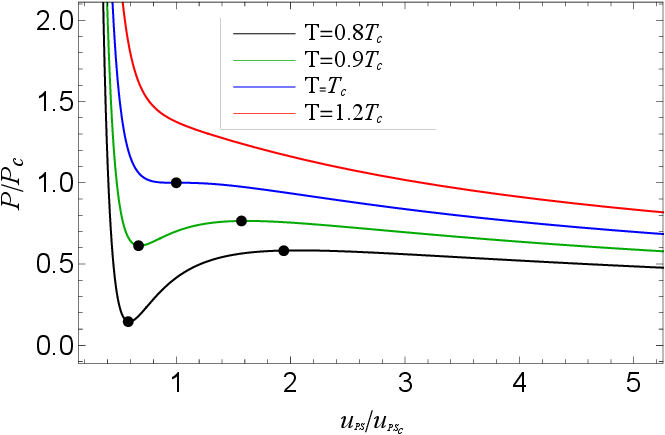}
			\end{tabular}
	\end{centering}
\caption{The behaviour of reduced pressure as a function of the radius of the photon sphere $r_{ps}/r_{psc}$ (\textit{left}) and critical impact parameter $u_{ps}/u_{psc}$ (\textit{right}) at different values of temperature. The  black dots  correspond to the extremal points in both plots.} \label{fig13}	
\end{figure*}
Solving these conditions, we can obtain the radius of unstable photon orbit  $r_{ps}$ and the critical impact parameter (or critical angular momentum) $u_{ps}$, for the unstable photon orbit. Substituting the expression of the effective potential from Eq.~(\ref{Veff1}), the above the conditions lead to
\begin{eqnarray}\label{c1}
 &&u_{ps}=\frac{a \left(r^3-2 l^2 M\right)+r l^2  \sqrt{a^2+r \left(\frac{r^3}{r^3-l^2}-2 M+r\right)}}{l^2 (2 M-r)}\Bigg|_{r=r_{ps}},
\end{eqnarray}
and the unstable photon orbit radius  $r_{ps}$ is obtained as  the largest real root of the following equation
\begin{eqnarray} \label{c2}
\frac{4 a \left(l^2 M+r^3\right)\left[-a+ \sqrt{a^2+r \left(\frac{r^3}{l^2}-2 M+r\right)}\right]+2 r (3 M-r) \left(l^2 (2 M-r)-r^3\right)}{l^2 r (r-2 M)+r^4}=0
\end{eqnarray}
A photon with a significant impact parameter $u$ will experience a small deflection angle. As $u$ approaches $u_{ps}$, the deflection angle will increase continuously and become infinitely large until $u_{ps}$ is reached. Light beams having an impact parameter less than $u_{ps}$ are absorbed by the black hole, while those with an impact parameter greater than $u_{ps}$ are deflected by the black hole (see Fig.~\ref{fig6a}).  Furthermore, we will conduct a qualitative analysis of the correlation between the phase change in the photon sphere's radius coordinate and the event horizon's radius coordinate. 
\subsection{Isobar and isotherm}
We aim to study the behaviours of $r_{ps}$ and $u_{ps}$ along the isobar and isotherm. It is important to find the critical values of unstable photon orbit radius and the critical impact parameter at the critical points of the phase transition given in Eq's~(\ref{criticalvaluesAdS}) and (\ref{criticalvaluesAdS2}). At the critical points of the phase transition,  the critical values were found to be
\begin{eqnarray}
 r_{psc}\approx 4.1092 a,\quad
 u_{psc}\approx5.81791 a.
\end{eqnarray}
By utilizing Equation (\ref{temp}), we can generate a graph that illustrates the relationship between the reduced temperature $(T/T_{C})$ and the reduced photon orbit radius ($r_{ps}/r_{psc}$) for various fixed values of pressure ($P/P_C$). This graph is depicted in Fig. \ref{fig12}. These isobaric curves demonstrate analogous patterns to the $T-r_{+}$ curves shown in Fig. \ref{fig7}. Based on the information provided in Fig. \ref{fig12}, it is evident that when the value of $P$ is less than $P_C$, a non-monotonic pattern emerges. This pattern consists of one local maximum and one minimum. This indicates that within a specific temperature range, there are two stable branches, representing LBH and SBH, respectively. Consequently, LBH-SBH phase transitions can occur. As $P$ increases, the two extremal points approach each other and merge at $P = P_C$, where a single reflection point is associated with the critical point. When the value of P is greater than the critical pressure ($P_C$), the curve exhibits only one local lowest point. Consequently, for a specific temperature, there is only one stable branch, preventing any possibility of a phase transition. In Fig. ~\ref{fig12}, we depict the isobaric $(T/T_{C})$ as a function of the reduced critical impact parameter ($u_{ps}/u_{psc}$). The plot exhibits comparable behaviours to the $T-r_{ps}$ curve. Based on this characteristic of the photon orbit radius and the crucial impact parameter (angular momentum) of the circular orbit, one can directly ascertain the occurrence of a thermodynamic phase transition by counting the number of extremal sites. These extremal points,  depicted as distinct black dots in Fig.~\ref{fig12} and  Fig.~\ref{fig13}.  Two extremal points characterize a first-order phase transition, while a second-order phase transition is characterized by only one extremal point. Nevertheless, the lack of a singular point does not necessarily indicate the absence of a phase transition. When the reduced pressure is plotted against the radius of the photon sphere, denoted as $r_{ps}/r_{psc}$, and the critical impact parameter, denoted as $u_{ps}/u_{psc}$, a similar result is obtained (see Fig.~\ref{fig13}). We note the following: i) When $\tilde{T}<1$, a thermodynamic first-order phase transition occurs; ii) At $\tilde{T}=1$, a second-order phase transition takes place; and iii) There is no phase transition for values of $\tilde{T}$ greater than 1.

\section{Conclusion}\label{sec5}
We explored the thermodynamic properties and phase transitions of a newly proposed gravitationally decoupled Kerr–AdS black hole, derived by interpreting the cosmological constant as vacuum energy. The metric for this black hole is more straightforward and exhibits richer geometric features than the standard Kerr-AdS solution, especially in how rotation manifests as a warped curvature.
First, we obtained an exact Kerr-AdS black hole using the gravitationally decoupled approach. The resulting Kerr-AdS black holes do not possess the old Kerr-AdS mathematical structure but exhibit numerous other fascinating properties. The line element for the new Kerr-AdS black holes \eqref{newKerrds} appears more straightforward than the standard Kerr-AdS \eqref{kdsstandard}, yet it encompasses a complex spacetime configuration. While the standard Kerr-AdS metric \eqref{kdsstandard} represents a $\Lambda$-vacuum solution, the metric for the new Kerr-AdS black holes \eqref{newKerrds} does not.

The impact of curvature $R$ is more substantial close to the rotating source, around $r\approx a$, and diminishes as you move farther away, where $R\approx 12/l^2$ for $r\gg a$. This phenomenon doesn't occur in the standard Kerr-AdS spacetime because it's designed to have a constant curvature throughout.  Unlike the standard Kerr-AdS black hole, the extremal black hole for the new Kerr-Ads occurs at a lower radius and value of $l$. We quantified this by explicitly solving the $\Delta(r)=0$ for real roots and demonstrating that for the black hole spin  $a \approx 0.8 M$, standard Kerr-AdS black hole and new Kerr-Ads, have extremal respectively, for $l = 13374$ with $r_+=0.62$ and $l = 0.75084$ with $r_+=0.55$. Interestingly, the $\Delta(r)=0$ of new Kerr-Ads spacetime can admit up to two real roots, viz., $r_2\; r_1$  corresponding to are Cauchy and event horizons with the possibility $r_2,\; = \; r_1$ for the extremal black holes. 

By interpreting the cosmological constant as a thermodynamic pressure and introducing its conjugate as the thermodynamic volume, we derived exact expressions for critical parameters such as mass, Hawking temperature, entropy, heat capacity, and Gibbs free energy.  We have investigated the thermodynamic criticality of newly formed gravitationally decoupled Kerr–AdS black holes, focusing on the $P-V$ diagram to uncover the intricate details of phase transitions. Our analysis extends to exploring the correlation between the photon orbit radius and these phase transitions, shedding light on how the geometric properties of black holes relate to their thermodynamic behaviour. This exploration is particularly significant as it suggests a profound connection between the gravitational characteristics of black holes, such as the behaviour of photon orbits, and their underlying thermodynamic processes.

We studied its thermodynamic phase transition and stability and then obtained a critical value at which the phase transition of the new Kerr-AdS black hole occurs. The behaviour of these black holes resembles that of a vdW fluid. Analyzing $\tilde{G}$ - $\tilde{T}$ indicates a first-order phase transition, while studying heat capacity points to a second-order phase transition. We have also investigated the correlation between the photon sphere radius and the thermodynamic phase transition in new Kerr-AdS black holes, which offers a fresh perspective on connecting the gravitational and thermodynamic aspects of black holes. Our findings reveal that the isobaric and isothermal curves related to the photon orbit radius ${r}_{ps}$, along with the minimum impact parameter ${u}_{ps}$, within the reduced parameter space, exhibit analogous thermal phase transition behaviours akin to conventional vdW-like fluids. Specifically, these curves in the ${T}-{r}_{ps}({u}_{ps})$ or ${P}-{r}_{ps}({u}_{ps})$ planes manifest oscillatory patterns, indicating a first-order phase transition characteristic of vdW fluids. The presence of oscillations signifies the existence of two extremes, and when these extremes coincide, a second-order phase transition occurs. Beyond the values corresponding to second-order phase transitions for other parameters, no such curves indicative of higher-order phases are observed.

In the reduced parameter space, the nonmonotonic behaviours of the photon sphere radius and minimum impact parameter below the critical pressure have emerged as valuable order parameters for identifying the phase transition between small and large black holes. Additionally, the application of the Maxwell equal area law has illustrated characteristic vdW-like oscillations in the $P-V$ diagram, signifying the phase transition and allowing for the derivation of fitting formulas for co-existence curves.

In conclusion, our study sheds light on the intricate dynamics of new Kerr-Ads black holes. It establishes connections between gravitational phenomena and thermodynamic principles, contributing significantly to the broader understanding of black hole thermodynamics and phase transitions. 

\section{Acknowledgement}
S.K would like to thank Council of Scientific and Industrial Re-search (CSIR), India for financial support through Senior Research Fellowship (Grant No. 09/466(0209)/2018-EMR-I). S.U.I would like to thank the University of KwaZulu-Natal and the NRF for the postdoctoral research fellowship. S.G.G. is supported by SERB-DST through project No. CRG/2021/005771. S.D.M acknowledges that this work is based upon research supported by the South African Research Chair Initiative of the Department of Science and Technology and the National Research Foundation.

\bibliography{KerrAdS}
\bibliographystyle{apsrev4-1}
\end{document}